\documentclass[aps,prxquantum,reprint,twocolumn,floatfix,groupedaddress,superscriptaddress]{revtex4-2}
\usepackage{pstricks,graphicx,amsmath,bbm,mathrsfs,amssymb,psfrag,pifont,times,mathptmx}
\usepackage[utf8]{inputenc}
\usepackage{hyperref}
\usepackage[english]{babel}
\usepackage{color}
\usepackage{ulem}
\usepackage{siunitx}
\usepackage{mathtools}
\usepackage{comment} 

\def\dag{{^{\dagger}}}

\def\ch{{\rm ch}}
\def\nn{{\rm n}}
\def\SB{{\rm SB}}
\def\EB{{\rm EB}}
\def\BM{{\rm BM}}
\def\GP{{\rm GP}}
\def\bmsigma{{\boldsymbol\sigma}}

\begin{document}
\title{Nonlocal continuous-variable gates by amplified optical connections}
\author{Michele N.~Notarnicola}
\email{michelenicola.notarnicola@upol.cz}
\affiliation{Department of Optics, Palack\'y University,
17. Listopadu 12, 779 00 Olomouc (Czech Republic) }

\author{Radim Filip}
\email{filip@optics.upol.cz}
\affiliation{Department of Optics, Palack\'y University,
17. Listopadu 12, 779 00 Olomouc (Czech Republic) }

\date{\today}
\begin{abstract}
Nonlocal quantum gates, coupling quantum systems located at a distance, are crucial for distributed quantum computing. To this aim, high-capacity optical noiseless connections between different processing units are essential for transmitting large amounts of information per mode. Simultaneously, optical quantum computing offers future high-speed multimode quantum processors. We propose a library of feasible protocols to implement a necessary nonlocal continuous-variable (CV) quantum nondemolition (QND) gate between two distant users sharing a quantum channel and exploiting classical communication. The users are endowed with a newly achieved high-fidelity and large-bandwith element - single-pass phase-sensitive optical parametric amplifier (OPA), that allows for both online squeezing and channel-loss compensation. 
The use of OPAs enhances quality of the resulting gate in terms of both excess noise and entangling capability. The proposed schemes are also applicable to CV cluster state fusion, providing a first step towards development of distributed CV measurement-based quantum computation.
\end{abstract}
\maketitle

\section*{Introduction}\label{sec:Intro}
Distributed quantum computing \cite{Cirac1999, Liang2007, Monroe2014, Li2024} is a crucial paradigm for the efficient scale up of quantum computers, obtained by mapping qubits to a nonlocal quantum network of smaller processing units
connected though quantum channels to build larger quantum processors. To this aim, the realization of nonlocal quantum gates that entangle different network nodes represents a fundamental step. Originally proposed in the early 2000s \cite{Gottesman1999, Eisert2000, Paternostro2003, Filip2004, Filip2005}, those gates allow two distant users sharing a quantum channel, Alice ($A$) and Bob ($B$), to effectively implement a joint unitary operation by means of both local operations and classical communication (LOCC) and the exchange of an ancillary system through the channel to mediate the interaction, hereafter referred to as mediator ($M$).
Experimental demonstrations at a distance have been only recently realized on discrete-variable (DV) platforms, where a nonlocal controlled-NOT (CNOT) operation has been distributed over distances of $0.6, 7, 10$ km, respectively \cite{Daiss2021, Liu2024, Iuliano2026}.

The continuous-variable (CV) analog of the CNOT gate is the quantum nondemolition (QND) coupling $U_{AB}=\exp(-i g x_A p_B/2)$, where $g \in \mathbb{R}$ is the gate gain, $x_j,p_j$, $j=A,B$, are quadrature operators, namely the position- and momentum-like variables, expressed in shot-noise units and satisfying $[x_j,p_k]= 2 i \delta_{jk}$ \cite{MyThesis}. Unlike beam splitter gates, the QND gate is advantageous for CV processing due to its versatility \cite{Kalajdzievski2021}. The input-output transformation in Heisenberg picture reads:
\begin{align}\label{eq:idealQND}
\begin{cases}
x_A'= x_A \,, \qquad  p_A'= p_A - g p_B \, ,  \\[1ex]
x_B'= x_B+ g x_A \,, \qquad p_B'= p_B \, ,
\end{cases}
\end{align}
showing $x_A$ and $p_B$ as nondemolishing variables, whose information is transmitted to $x_B'$ and $p_A'$, respectively. Local QND interaction is directly available in quantum optomechanics and atomic ensembles \cite{Manukhova2020, Thomas2021, Manukhova2022, Manukhova2024,Lei2025}, and its all-optical implementation, based on measurement-induced amplifiers with offline squeezers, has also been tested \cite{Yokoyama2014, Shiozawa2018}. However, use of online squeezers instead of measurement-induced schemes \cite{Filip2005} would be essential for the realization of nonlocal CV gates.

The recent development of single-pass optical parametric amplifiers (OPAs) \cite{OPA2019, OPA2020, OPA2020d, OPA2021,OPA2023} shed new light on this subject. Waveguide-based high-fidelity OPAs are $\chi^{(2)}$-fabricated waveguides that perform online broadband squeezing of optical signals throughout propagation, showing higher optical nonlinearities than conventional cavity-based methods \cite{Notarnicola2022, OPO2024a,OPO2024b}. As active online elements, they may be employed to both build QND interaction in optics \cite{Shiozawa2018} and perform signal restoration as phase-sensitive amplifiers (PSAs) \cite{MyThesis, Notarnicola2024, Yan2022, Zhao2022}, offering new possibilities for distributed quantum information. Moreover, OPAs have been also investigated for quantum state tomography \cite{Shaked2018, Frascella2021Lett, Kalash2023,Racz2024}, certification of quantum non-Gaussianity \cite{Kalash2025}, multiplexed quantum key distribution \cite{Eldan2026} and quantum interferometry \cite{Natan2026}.

\begin{figure*}
\includegraphics[width=0.99\linewidth]{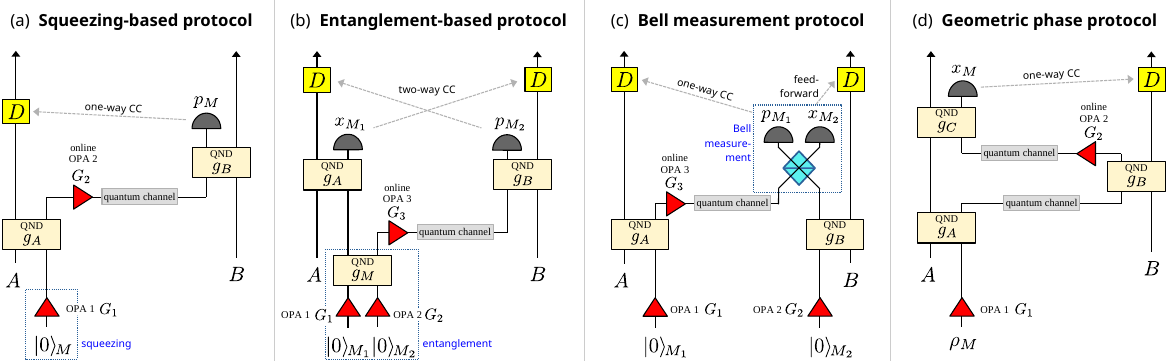}
\caption{Schemes for nonlocal QND gate implementation with different resources, tolopogy and applications. The squeezing-based ($\SB$) protocol (a) uses an offline-squeezed vacuum mediator with single quantum channel and one-way classical communication (CC). The entanglement-based ($\EB$) scheme (b) exploits advantageous entanglement pre-sharing, useful for entanglement distillation protocols, by designing an optimized two-mode state that fits into QNDs $g_{A(B)}$, at the cost of two-way CC, while the Bell measurement ($\BM$) protocol (c) replaces the prior resource distribution with online entanglement generation by Bell measurement, being suitable for cluster-state manipulation, and reducing the classical resources to one-way CC and feed-forward. Unlike the previous setups, the geometric phase ($\GP$) protocol enables nonlocal gates independent of the mediator momentum-like quadrature noise, at the costs of double use of quantum channel and one more gate at Alice's side. As discussed in the main text, the use of a further online OPA to assist the first channel does not bring advantage. In all setups, it is also possible to design local Gaussian gates that integrate QNDs together with the OPAs.  Displacements might also be done by QND interaction with classical coherent light controlled by the measured results. In all schemes we further assume ideal homodyne detection, for if detectors have low quantum efficiency, one can use a further OPA as pre-amplifier \cite{Leonhardt1994}.}\label{fig:01-Protocols}
\end{figure*}

In this work, we propose a list of different measurement-induced OPA-assisted protocols to distribute essential QND gates at a distance, assuming QND~(\ref{eq:idealQND}) is locally available to both the remote users Alice and Bob, and employing mediators that exhibit different quantum resources: squeezing [Fig.~\ref{fig:01-Protocols}(a)], entanglement [Fig.~\ref{fig:01-Protocols}(b)], Bell measurement [Fig.~\ref{fig:01-Protocols}(c)] and the geometric phase effect by sequential local QNDs [Fig.~\ref{fig:01-Protocols}(d)]. We characterize the resulting nonlocal entangling gates in terms of both Gaussian logarithmic negativity and excess noise
and optimize their performance. We prove the first three schemes to be equivalent with one another, thus offering flexibility for different implementations, whilst the geometric phase protocol to give a factor $2$ improvement over the others in both the limits of small and large losses. In all cases, online OPAs with optimized gains prove useful to protect the information carried by the mediator against the channel losses, and are crucial to preserve entanglement at large loss. This is especially relevant for the first experimental demonstrations of the nonlocal gates.
Further, we demonstrate application of our approach for fusion of distant cluster states \cite{Yokoyama2013,Schwartz2016, Larsen2019} into a distributed network \cite{Xin2023}. 

%

\section*{Results}

\subsection*{A model for realistic waveguide OPAs}\label{sec:OPA}
Together with online squeezing, a realistic waveguide OPA is expected to have distributed losses occurring throughout propagation along the waveguide, beyond incoupling and outcoupling losses. Accordingly, the optical output state is described by the master equation $\dot{\rho}= -i [H_{\rm OPA}, \rho] + \gamma \, {\cal L}[a]  \rho$, where $H_{\rm OPA}= i \chi [ (a\dag)^2-a^2]/2$ is the squeezing Hamiltonian, $\chi \in \mathbb{R}$, $\gamma\ge0$ is the linear loss rate and ${\cal L}[O]\rho= O \rho O\dag - \{O\dag O, \rho\}/2$ is the corresponding Lindblad operator.
An equivalent description for the quadratures $x=a+a\dag$ and $p=i(a\dag-a)$, in the Heisenberg picture, derived in Methods, is given in terms of pure squeezing followed by squeezed-thermal-loss channel:
\begin{align}\label{eq:OPAModel}
x'= \sqrt{\eta G} \, x + \sqrt{1-\eta} \, x_\nn \,, \quad 
p'= \sqrt{\frac{\eta}{G}}\, p + \sqrt{1-\eta} \, p_\nn \, ,
\end{align}
where: 
\begin{align}\label{eq:ParametersOPA}
G= e^{2\chi t}>0 \qquad \mbox{and} \qquad \eta=e^{-\gamma t} \ge 0
\end{align}
are the amplifier gain and efficiency, respectively, $t$ being the interaction time fixed by the waveguide length, while $x_\nn,p_\nn$ are quadrature operators of the channel-noise mode, excited in a squeezed thermal state with $\langle x_\nn\rangle=\langle p_\nn \rangle=0$ and
\begin{align}\label{eq:xnnpnn}
\langle x_\nn^2\rangle= \frac{1-\eta G}{1-\eta} \, \frac{\ln(\eta)}{\ln(\eta G)} \,, \quad
\langle p_\nn^2\rangle= \frac{1-\eta/G}{1-\eta} \, \frac{\ln(\eta)}{\ln(\eta/G)} \, .
\end{align}
For a lossless amplifier $\eta=1$, we retrieve the unitary squeezing transformation $x'= \sqrt{G} \, x $ and $p'= p/\sqrt{G}$, that amplifies (squeezes) quadrature $x$ if $G>1$ ($G<1$), whereas in the absence of squeezing, $G=1$, Eq.~(\ref{eq:OPAModel}) yields a pure-loss channel. 
Advantageously, this representation is also easy to combine with the incoupling and outcoupling loss to an unified operator transformations. Nevertheless, while coupling losses are described by pure-loss channels, distributed OPA losses introduce non-negligible external thermal noise to the signal mode, as even for $1-\eta\ll 1$ we have $\langle x_\nn^2\rangle \approx (G-1)/\log(G)$ and $\langle p_\nn^2\rangle \approx (G^{-1}-1)/\log(G^{-1})$, with $\langle x_n^2\rangle$ rapidly growing with gain $G$, which has potentially more crucial impact on realistic applications.
For the sake of completeness, the role of imperfect incoupling procedures is further discussed in Methods.

\subsection*{Nonlocal QND gates}\label{sec:Gates}

Fig.~\ref{fig:01-Protocols} schematizes the proposed protocols for distributed gates, where Alice and Bob wish to implement a nonlocal QND gate with target gain $g>0$.
The first minimal scheme, named squeezing-based ($\SB$) protocol [Fig.~\ref{fig:01-Protocols}(a)], requires two local QNDs, and both a single quantum and classical-communication channels \cite{Filip2004}. The quantum channel is a pure-loss channel of transmissivity $T\le 1$. We assume Alice to possess an auxiliary vacuum mediator $M$, being offline squeezed by an OPA of parameters $\eta_1,G_1$, and then coupled to $A$ by local QND~(\ref{eq:idealQND}) of gain $g_A>0$. Thereafter, Alice transmits $M$ to Bob, through the pure-loss channel, preceded by a second OPA with $\eta_2,G_2$, operating as PSA to pre-compensate for channel losses \cite{MyThesis, Notarnicola2024}. Crucially, the PSA should be placed before the channel, so as to amplify only $M$ and not the additional channel-loss mode. Bob then couples $M$ to $B$ by a second QND gate of gain $g_B= g/(g_A \sqrt{\eta_2 G_2 T})$, measures quadrature $p_M$ via homodyne detection and classically communicate the measurement outcome $\bar{p}$ to Alice, who applies the momentum-like displacement $p_A \to p_A +g_A \sqrt{\eta_2 G_2 T} \bar{p}$.

The second protocol is the entanglement-based ($\EB$) [Fig.~\ref{fig:01-Protocols}(b)], where, now, Alice and Bob share an entangled resource state created from two vacuum mediators $M_{1(2)}$ squeezed by OPAs 1(2) and undergoing local QND of gain $g_M\equiv-g_0/g_A<0$. As discussed in \cite{vanLoock2007}, the state might be alternatively obtained by two squeezers and beam-splitter interaction. The resource, located in Alice's lab, can be advantageously priorly distributed between $A$ and $B$ through the channel, with OPA 3 acting as PSA, and stored in quantum memories before its use. On their own branch, Alice and Bob implement QNDs of gains $g_A>0$ and $g_B= g/(g_A \sqrt{\eta_3 G_3 T})$, respectively, followed by homodynes of $x_{M_1}$ and $p_{M_2}$. After two-way classical communication of the outcomes $\bar{x},\bar{p}$, displacements $p_A \to p_A +g_0 \sqrt{\eta_3 G_3 T} \bar{p}$ and $x_B \to x_B +g \bar{x}/g_A$ are performed. Here, the possibility to pre-share entanglement makes the $\EB$ configuration suitable for (non-Gaussian) entanglement distillation schemes \cite{EntanglementDist1,EntanglementDist2}, 
at the additional cost of 
a two-way communication channel, which can actually be one physical channel used in both directions.


A similar topology is the Bell measurement ($\BM$) scheme [Fig.~\ref{fig:01-Protocols}(c)], that replaces the a priori entangled-state distribution of $\EB$ with online entanglement generation by Bell measurement and only one-way CC. Here, Alice and Bob first perform local QNDs of gain $g_{A(B)}>0$ with two independent vacuum mediators $M_{1(2)}$ priorly squeezed by OPAs 1(2), after which $A$ sends $M_1$ to $B$ via the channel, with OPA 3 as PSA. The gain of Bob's QND is chosen as $g_B= g/(g_A \sqrt{\eta_3 G_3 T})$. Then, in Bob's lab $M_1$ and $M_2$ are entangled by Bell measurement, realized by a balanced beam splitter and homodynes of $p_{M_1}$ and $x_{M_2}$. Given the outcomes $\bar{x},\bar{p}$, we finally implement displacements $p_A \to p_A +g_A \sqrt{2\eta_3 G_3 T} \bar{p}$ and $x_B \to x_B -\sqrt{2}g_B \bar{x}$, that require one-way CC $B\to A$ and feed-forward on Bob's side. We also note that an equivalent setup may be obtained by considered an active QND-based Bell measurement \cite{Lades2005}, where the balanced beam splitter is replaced by a further QND gate of gain $g_M=- g/(g_A g_B\sqrt{ \eta_3 G_3 T})<0$, while Bob's gain $g_B$ is now considered as a free parameter.
Unlike $\SB$, in the $\BM$ protocol Alice and Bob are equally treated, therefore the QND interactions with gains $g_{A(B)}$ may be part of more complex local operations before the nonlocal QND gate is implemented. Moreover, compared to the $\EB$, the $\BM$ strategy does not require prior entanglement distribution, thus providing a useful topology for cluster-state manipulation \cite{Yokoyama2013,Schwartz2016, Larsen2019}.

For all these three schemes $\rm s= \SB, \EB, \BM$, the overall modes transformation for $A, B$ (see Methods for details) is obtained as:
\begin{align}\label{eq:NonLocalTrans}
\begin{cases}
x_A'= x_A \,, \qquad  p_A'= p_A - g p_B + {\cal N}_p^{(\rm s)} \, ,  \\[1ex]
x_B'= x_B+ g x_A + {\cal N}_x^{(\rm s)}\,, \qquad p_B'= p_B \, ,
\end{cases}
\end{align}
implementing an additive-Gaussian-noise map that effectively realizes nonlocal unity-gain QND~(\ref{eq:idealQND}) between $A$ and $B$ up to additive Gaussian noise modes ${\cal N}_{x(p)}^{(\rm s)}$, that collect the mediator, channel and OPA noise contributions. We have:
\begin{subequations}\label{eq:NoiseSB}
\begin{align}
&{\cal N}_p^{(\SB)}= - g_A  (1- \eta_2 T) \left( \sqrt{\frac{\eta_1}{G_1}} p_M + \sqrt{1-\eta_1} p_{\nn ,1} \right) \nonumber \\
&\hspace{.5cm}+ g_A \sqrt{  \eta_2 G_2 T} \left( \sqrt{1-T} p_\ch + \sqrt{(1-\eta_2)T} p_{\nn,2} \right) \,,
\end{align}
\begin{align}
&{\cal N}_x^{(\SB)}=\frac{g}{g_A}  \left( \sqrt{\eta_1 G_1} x_M + \sqrt{1-\eta_1} x_{\nn ,1} \right)  \nonumber \\
&\hspace{.5cm}+ \frac{g}{g_A \sqrt{ \eta_2 G_2 T}} \left( \sqrt{1-T} x_\ch + \sqrt{(1-\eta_2)T} x_{\nn,2} \right) \, ,
\end{align}
\end{subequations}
 for the $\SB$ protocol, while for $\EB$ and $\BM$:
\begin{subequations}\label{eq:NoiseEB}
\begin{align}
&{\cal N}_p^{(\EB)}= - g_0  (1- \eta_3 T) \left( \sqrt{\frac{\eta_2}{G_2}} p_{M_2} + \sqrt{1-\eta_2} p_{\nn ,2} \right) \nonumber \\
&\hspace{.5cm}+ g_0 \sqrt{\eta_3 G_3 T} \left( \sqrt{1-T} p_\ch + \sqrt{(1-\eta_3)T} p_{\nn,3} \right)  \nonumber \\
&\hspace{.5cm}- g_A  \left( \sqrt{\frac{\eta_1}{G_1}} p_{M_1} + \sqrt{1-\eta_1} p_{\nn ,1} \right)\,,
\end{align}
\begin{align}
&{\cal N}_x^{(\EB)}=\frac{g}{g_0}  \left( \sqrt{\eta_2 G_2} x_{M_2} + \sqrt{1-\eta_2} x_{\nn ,2} \right)  \nonumber \\
&\hspace{.5cm}+ \frac{g}{g_0 \sqrt{\eta_3 G_3 T}} \left( \sqrt{1-T} x_\ch + \sqrt{(1-\eta_3)T} x_{\nn,3} \right) \, ,
\end{align}
\end{subequations}
and
\begin{subequations}\label{eq:NoiseBM}
\begin{align}
&{\cal N}_p^{(\BM)}=  - g_A  (1- \eta_3 T) \left( \sqrt{\frac{\eta_1}{G_1}} p_{M_1} + \sqrt{1-\eta_1} p_{\nn ,1} \right) \nonumber \\
&\hspace{.5cm}+ g_A \sqrt{\eta_3 G_3 T} \left( \sqrt{1-T} p_\ch + \sqrt{(1-\eta_3)T} p_{\nn,3} \right)   \nonumber \\
 &\hspace{.5cm} +g_A \sqrt{\eta_3 G_3 T} \left( \sqrt{\frac{\eta_2}{G_2}} p_{M_2} + \sqrt{1-\eta_2} p_{\nn ,2} \right)\,,
\end{align}
\begin{align}
&{\cal N}_x^{(\BM)}=\frac{g}{g_A}  \left( \sqrt{\eta_1 G_1} x_{M_1} + \sqrt{1-\eta_1} x_{\nn ,1} \right)  \nonumber \\
&\hspace{.5cm}+ \frac{g}{g_A \sqrt{ \eta_3 G_3 T}} \left( \sqrt{1-T} x_\ch + \sqrt{(1-\eta_3)T} x_{\nn,3} \right) \, ,
\end{align}
\end{subequations}
where $x_\ch, p_\ch$ are the channel noise quadratures, assumed in the vacuum. As we can see, the $\EB$ protocol in the limits $g_A\ll 1$ and $G_1\gg1$, for which $g_A^2 [\eta_1/G_1 + (1-\eta_1) \langle p_{\nn,1}^2 \rangle ] \ll 1$, becomes independent of mediator $M_1$ and turns out to be equivalent to $\SB$. The same happens for the $\BM$ scheme that, when 
$G_2\gg1$, is independent of $M_2$ as $[\eta_2/G_2 + (1-\eta_2) \langle p_{\nn,2}^2 \rangle ] \ll 1$,so that all three schemes eventually yield the same performance, ${\cal N}_{x(p)}^{(\EB)} \approx {\cal N}_{x(p)}^{(\BM)} \approx {\cal N}_{x(p)}^{(\SB)}$.

Instead, a different result is provided by the last protocol in Fig.~\ref{fig:01-Protocols}(d), named geometric phase ($\GP$), as being inspired on the quantum geometric phase effect \cite{Kupcik2015} observed in quantum transducers \cite{Bagci2014, Jockel2015, Vostrosablin2017, Vostrosablin2018, Zeuthen2020}. It uses a single-mode mediator $M$ excited in arbitrary quantum state $\rho_M$. Additionally, the scheme employs squeezing by OPA~1, but, unlike the $\SB$, it requires three gates and two quantum channel uses. First, Alice implements QND of gain $g_A>0$ with $M$. Then, she transmits $M$ to Bob, who couples performs QND of gain $g_B>0$ and sends it back to Alice, now using OPA 2 as PSA before the second passage into the channel. Finally, Alice implements another QND of gain $g_C=-g \sqrt{G_2/(\eta_2 T g_B^2)}<0$, measures $x_M$ and communicate the outcome $\bar{x}$ to Bob, who displaces $x_B \to x_B +g_B \Gamma \, \bar{x}$, with 
$$\Gamma= \frac{g- g_A g_B \sqrt{T}}{- g\sqrt{G_2/(\eta_2 T)} + g_A g_B T \sqrt{\eta_2 G_2} } \, .$$ For completeness, we checked that inserting a further OPA to assist also the first channel does not interestingly bring advantage, which simplifies the protocol. If lossless ($\eta_1=1$), OPA~1 might be equivalently moved after the first QND and before the channel, while in the realistic case $\eta_1<1$ this would degrade the gate entangling power, due to the external noise introduced by the amplifier.
Eventually, the protocol implements the additive-Gaussian-noise map of Eq.~(\ref{eq:NonLocalTrans}) with the noise modes:
%
%
\begin{subequations}\label{eq:NoiseGP}
\begin{align}
{\cal N}_p^{(\GP)}&= - \left(g_A - \frac{g \sqrt{T}}{g_B}\right) \left( \sqrt{\frac{\eta_1}{G_1}} p_M + \sqrt{1-\eta_1} p_{\nn ,1} \right)\nonumber \\
&\hspace{.5cm}+ \frac{g \sqrt{1-T}}{g_B} \left( p_{\ch,1} + \sqrt{\frac{G_2}{\eta_2 T}} p_{\ch,2}\right)  \nonumber \\
&\hspace{.5cm}+ \frac{g}{g_B} \sqrt{\frac{G_2(1-\eta_2)}{\eta_2}} p_{\nn,2} \, ,
\end{align}
\begin{align}
{\cal N}_x^{(\GP)}&= g_B \left(1+ \Gamma \sqrt{\eta_2 G_2 T} \right)  \Bigg[ \sqrt{\eta_1 G_1 T} x_M \nonumber \\
&\hspace{.5cm} +  \sqrt{(1-\eta_1)T} x_{\nn ,1} + \sqrt{1-T} x_{\ch,1} \Bigg]  \nonumber \\
&\hspace{.5cm}+ g_B \Gamma \left( \sqrt{1-T} x_{\ch,2} + \sqrt{(1-\eta_2)T} x_{\nn,2} \right) \, .
\end{align}
\end{subequations}
In particular, for lossless channels ($T=1$) and ideal OPAs ($\eta_{1(2)}=1$), the protocol becomes fully independent of the mediator state $\rho_M$ in the limit $g_Ag_B=g$, thus implementing the target gate even in the presence of a noisy mediator, e.g. in a highly thermal state \cite{Kupcik2015}. Conversely, when $T<1$ we only get rid of quadrature $p_M$ by condition $g_Ag_B=g\sqrt{T}$, whereas the mediator noise in quadrature $x_M$ can be suppressed by use of large OPA~1 squeezing, leading to conditions
\begin{align}\label{eq:ConditionGP}
g_A= \frac{g \sqrt{T}}{g_B} \qquad \mbox{and} \qquad G_1 \to 0 \, .
\end{align}
In conclusion, the $\GP$ scheme allows for efficient nonlocal QND gates, inherently independent of the momentum-like mediator noise, at the cost of large offline squeezing and double use of the quantum channel. The impact of finite squeezing $G_1>0$ is discussed in Methods. In light of this, the principle behind $\GP$ is different from the previous protocols, so that even a combination of the schemes in Fig.~\ref{fig:01-Protocols}(b) and (c) would yield worse performance.

\begin{figure}
\includegraphics[width=0.95\columnwidth]{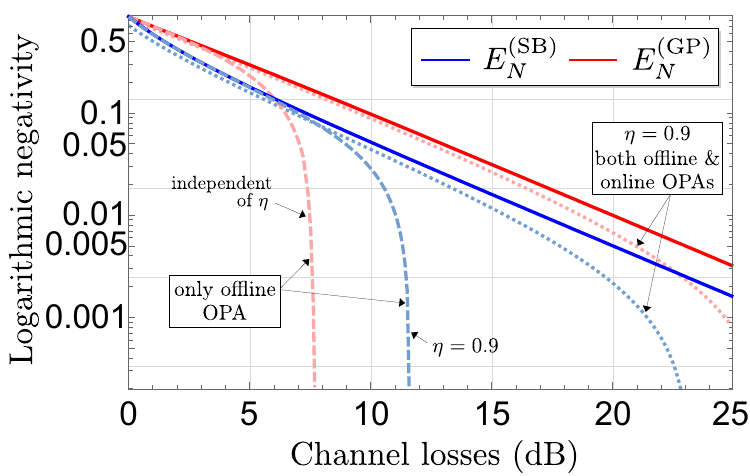}
\caption{Logaritmic negativity $E_N^{(\rm s)}$, ${\rm s}=\SB,\GP$, for $\eta=1$ (solid lines) and target gain $g=1$ as a function of the channel losses, equal to $-10\log_{10} (T)$ dB. Both the schemes allow for arbitrary long-distance entanglement for $\eta=1$, and $\GP$ outperforms $\SB$, with a factor $2$ advantage for $1-T\ll 1$, where $\partial_T E_N^{(\GP)}=\partial_T E_N^{(\SB)}/2$. Light-colored lines are logarithmic negativities $E_{N,{\rm off(on)}}^{(\rm s)}$ with realistic OPAs ($\eta=0.9$) for the two cases of only offline squeezing and both online and offline OPAs, that both induce a maximum transmission loss. For $1-T\ll 1$, online OPA is useless for the $\SB$ and useful for the $\GP$, provided that $\eta >\eta_\GP(T)$. The $\EB$ and $\BM$ protocols are equivalent to the $\SB$, but have advantages based on distributed prior entanglement or posterior Bell measurements.}\label{fig:02-GatesNeg}
\end{figure}

\begin{figure*}
\includegraphics[width=0.99\linewidth]{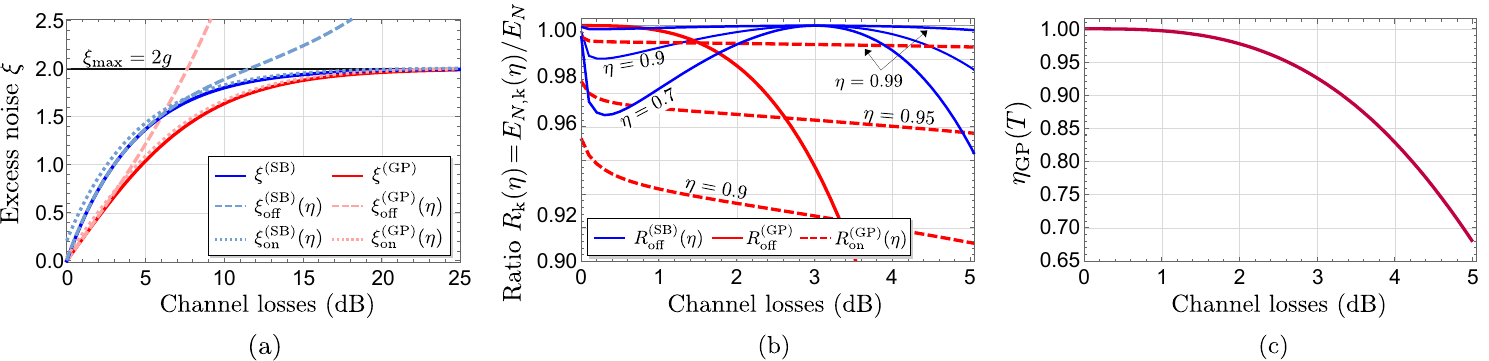}
\caption{(a) Excess noise $\xi^{(\rm s)}$, ${\rm s}=\SB,\GP$, for $\eta=1$ (solid lines) and target gain $g=1$ as a function of the channel losses. For $1-T\ll 1$, $\GP$ gives a factor~$2$ advantage, as $\xi^{(\GP)}=2g(1-T)/(1+T) \approx g (1-T) =\xi^{(\SB)}/2$. Light-colored lines refer to the cases of lossy OPAs with $\eta=0.9$. Entanglement is lost whenever $\xi^{(\rm s)}$ beats the maximum tolerable noise $\xi_{\rm max}=2g$. (b) Entanglement ratio $R^{(\rm s)}_{\rm k}=E_{N,{\rm k}}^{(\rm s)}(\eta)/E_{N}^{(\rm s)}$, k=off,on, between the realistic case of noisy amplifiers and the ideal OPA scheme for $g=1$ and different efficiencies $\eta$. When channel loss $1-T\ll 1$, online OPA is useless for the $\SB$ scheme as $\xi_{\rm on}^{(\SB)}(\eta)$ (dotted blue line) is larger than $\xi_{\rm off}^{(\SB)}(\eta)$ (dashed blue line), whereas, in the $\GP$ protocol, noisy online amplification helps provided that the OPA efficiency is $\eta>\eta_{\GP}(T)$. The $\EB$ and $\BM$ protocols are equivalent to the $\SB$, but have advantages based on distributed prior entanglement or posterior Bell measurements. (c) Threshold OPA efficiency $\eta_{\GP}(T)$ as a function of the channel losses.}\label{fig:03-GatesNoise}
\end{figure*}

To assess the quality of the built unity-gain gates~(\ref{eq:NonLocalTrans}), we consider as figures of merit (i) the quadrature excess noise $\xi^{(\rm s)} ={\rm Var}[ {\cal N}_x^{(\rm s)}] ={\rm Var} [{\cal N}_p^{(\rm s)}]$, $\rm s= \SB, \EB, \BM, \GP$, in the case of symmetric noise on both Alice and Bob sides, that maximizes the amount of output entanglement and, consequently, (ii) the Gaussian logarithmic negativity in such case 
assuming $A,B$ initially in the vacuum, 
derived in Methods as:
\begin{align}\label{eq:LN}
E_N^{(\rm s)}= -\frac12 \ln \left(1+2g^2+\xi^{(\rm s)} - 2g \sqrt{1+g^2+\xi^{(\rm s)}} \right) 
\end{align}
for $\xi^{(\rm s)}<2g$ and 0 otherwise.
The local QND and OPA gains are optimized to minimize (i), thus maximizing (ii). Given that the $\SB,\EB,\BM$ protocols are equivalent with one another, we limit ourselves to compare to the $\SB$ and $\GP$ schemes, depicted in Fig.s~\ref{fig:01-Protocols}(a) and (d), respectively.

For ideal OPAs with $\eta=1$, the optimal configuration for the $\SB$ scheme is obtained by the choices
\begin{align}\label{eq:ConditionSB}
G_1 G_2 = \frac{1-T}{T} \qquad \mbox{and} \qquad g_A= \sqrt{\frac{g \, G_1}{1-T}} \, ,
\end{align}
for which the excess noise reads $\xi^{(\SB)}= 2g(1-T)$. The combined effect of the two OPAs gives flexibility for the implementation, as one can either prepare the squeezing resource to the gain $G_1=(1-T)/T$ and remove online OPA~2, or alternatively, if a squeezed mediator with fixed $G_1$ is available, tune OPA~2 to the gain $G_2=(1-T)/(G_1 T)$ to meet condition~(\ref{eq:ConditionSB}). In this way, information about $x_A$ to be transferred to Bob, carried by $x_M$ after the first QND, is protected against the channel losses,
while that on $p_B$ is passed to Alice by the final conditional displacement.
Remarkably, the maximum entangling power is obtained for finite OPA gains, $G_1 G_2 <\infty$, and the corresponding value of $E_N^{(\SB)}$ equals that of the scheme proposed in \cite{Filip2004}, where an infinitely squeezed mediator was considered as resource, and the channel was assisted by phase-insensitive amplifiers.
Consistently with the previous considerations, the $\EB$ and $\BM$ protocols yields the same performance of the $\SB$ in the limits $g_A^2/G_1 \ll 1$, $G_2 G_3 = (1-T)/T$, $g_0= \sqrt{g G_2/(1-T)}$ ($\EB$) and 
$G_2 \gg 1$, $G_1 G_3 = (1-T)/T$, $g_A= \sqrt{g G_1/(1-T)}$ ($\BM$), and provide efficient methods applicable to the different use cases.

On the contrary, the $\GP$ logarithmic negativity $E_N^{(\GP)}$ is maximized for $G_2=T$ and $g_B= \sqrt{g(1+T)}$, together with the asymptotic mediator-independence conditions~(\ref{eq:ConditionGP}), leading to $\xi^{(\GP)} = 2g (1-T)/(1+T) < \xi^{(\SB)}$ and $E_N^{(\GP)}>E_N^{(\SB)}$. Now, OPA~2 perfectly compensates the losses of the second channel on the $p_M$ quadrature that carries information about $p_B$ to Alice, at the expense of reducing that on $x_A$, which is eventually restored by Bob after displacement. Moreover, the large anti-squeezing noise of $p_M$ is removed by the last QND gate at Alice's side thanks to the geometric phase trick.
As shown in Fig.~\ref{fig:02-GatesNeg} (solid lines), both $\SB$ and $\GP$ schemes yield nonzero negativity for all $T\le 1$, guaranteeing entanglement at arbitrary large losses. In particular, while the optimum performance for the $\SB$ is achieved by the interplay between OPAs~1(2) with gains $G_{1(2)}$, in the $\GP$, online OPA~2 is crucial for long-distance entanglement, as using only offline OPA~1 would make $E_N^{(\GP)}$ vanish at large losses (dashed red line in Fig.~\ref{fig:02-GatesNeg}). Moreover, for small losses $1-T\ll1$, being of interest for real implementations, we find:
\begin{align}\label{eq:ConditionSB}
E_N^{(\rm s)} \approx E_N^{(0)} - \gamma_{\rm s} (1-T) \, , \quad {\rm s}= \SB,\GP\, ,
\end{align}
with $E_N^{(0)}$ computed from~(\ref{eq:LN}) with $\xi=0$, $\gamma_{\SB}=g (1-g/\sqrt{1+g^2})/(1+2 g^2-2g \sqrt{1+g^2})$ and $\gamma_\GP= \gamma_\SB/2$, whilst in the long-distance regime $T\ll 1$, $E_N^{(\SB)}\approx g T/(1+g)$ and $E_N^{(\GP)}\approx 2 E_N^{(\SB)}$, proving $\GP$ to introduce a factor $2$ improvement over $\SB$ in both the limits, see the solid lines of Fig.~\ref{fig:02-GatesNeg}.

The scenario changes with realistic OPAs of non-unit efficiency $\eta<1$. Since the online amplifiers introduce thermal noise even for $1-\eta \ll 1$, we identify two different scenarios, case (off): only offline squeezing produced by OPA 1 is considered, while the online PSAs are removed, corresponding to the choices $\eta_1=\eta<1$, $\eta_k=G_k=1$ for $k>1$ in Fig.~\ref{fig:01-Protocols}; case (on): we keep all online and offline lossy OPAs with $\eta_k=\eta<1$. The corresponding negativities $E_{N,{\rm off(on)}}^{(\rm s)}(\eta)$, ${\rm s}=\SB,\GP$, depicted in Fig.~\ref{fig:02-GatesNeg}, exhibit a maximum transmission loss, after which they drop to $0$, preventing long-distance entanglement. 
The entanglement break is a consequence of the thermal noise introduced by the lossy OPAs, that makes the gate excess noise $\xi_{\rm off(on)}^{(\rm s)}(\eta)$, reported in Fig.~\ref{fig:03-GatesNoise}(a), exceed the maximum tolerable threshold $\xi_{\rm max}=2g$ for large losses, so that Eq.~(\ref{eq:LN}) vanishes. 

Furthermore, for lossless channel ($T=1$) we observe a gap between $E_{N,{\rm on}}^{(\rm s)}(\eta)$ (dotted lines in Fig.~\ref{fig:02-GatesNeg}) and the protocols employing ideal OPAs (solid lines), being absent in the (off) case (dashed lines in Fig.~\ref{fig:02-GatesNeg}), which is direct consequence of the internal losses of online OPAs. In fact, when $T=1$, the optimal online squeezing is low, $G_2 \approx 1$, as discussed in App.~\ref{app:NONLOCALAPP}, so that the online OPA effectively acts as a lossy channel with transmissivity $\eta$, see~(\ref{eq:OPAModel}), and reduces the entangling capability of the gate. On the contrary, when $T\ll 1$, $G_2$ gets large values, thus counteracting the detriments of the internal losses. In such regime, we observe a crossing between $E_{N,{\rm off}}^{(\rm s)}(\eta)$ and $E_{N,{\rm on}}^{(\rm s)}(\eta)$ and online OPAs, although noisy, become helpful to considerably increase the maximum transmission loss.
Instead, in the opposite limit $1-T\ll 1$ the $\SB$ and $\GP$ schemes behave differently. In the former, 
the best strategy is to use only OPA 1 as offline squeezer, since for $1-\eta\ll 1$ the excess noise in case (off) reads:
\begin{align}\label{eq:ExSBeta}
\xi_{\rm off}^{(\SB)}(\eta) &\approx \xi^{(\SB)}+ g (1-\eta) \left[\frac{1-2 T}{2 T \log \left(\frac{1-T}{T}\right)}-(1-T) \right] \, ,
\end{align}
while in case (on) we find $\xi_{\rm on}^{(\SB)}(\eta) \approx \xi_{\rm off}^{(\SB)}(\eta) + 2g(1-\eta)$ for both $1-\eta, 1-T\ll 1$. Therefore, for small channel losses the introduction of online OPA 2, even if slightly lossy, would reduce the gate entangling power by adding noise. In such case, the amplification gain would be masked by the additional internal losses, whilst in the opposite regime $T\ll 1$ the OPA effect remains helpful.
The $\SB$ output entanglement is also robust to the reduced efficiency, as emerges from the ratios $R^{(\rm s)}_{\rm k}(\eta)=E_{N,{\rm k}}^{(\rm s)}(\eta)/E_{N}^{(\rm s)}$, k=off,on, plotted in Fig.~\ref{fig:03-GatesNoise}(b). In fact, $R^{(\SB)}_{\rm off}(\eta)$ has a non-monotonic behavior (blue lines), that for $1-T\ll 1$ decreases to a minimum $\approx 99.9, 99, 97.3 \%$ for $g=1$ and $\eta=0.99,0.9, 0.7$, respectively, and then re-approaches 1 at $3$ dB losses, i.e. $T=1/2$, thus showing limited entanglement decay w.r.t. the ideal case.
On the other hand, in the $\GP$ protocol lossy online OPA 2 is still beneficial when $1-T\ll 1$, provided its internal losses to be small enough. Indeed, the excess noise for case (off) is independent of $\eta$, due to the large offline squeezing $G_1 \to 0$, and equal to:
\begin{align}
\xi_{\rm off}^{(\GP)} = g \frac{1-T}{\sqrt{T}} > \xi^{(\GP)} \, ,
\end{align}
whereas, for $1-\eta \ll 1$, $\xi_{\rm on}^{(\GP)}$ is expanded as:
\begin{align}
\xi_{\rm on}^{(\GP)} (\eta) &\approx \xi^{(\GP)}+ g (1-\eta)\left[\frac{T (1-T)}{(1+T)^2}-\frac{1-T}{2 \log T} \right] \, ,
\end{align}
so that online OPA 2 is helpful if $\xi_{\rm on}^{(\GP)} (\eta)<\xi_{\rm off}^{(\GP)}$, namely if $\eta$ is larger than the threshold function $\eta_{\GP}(T)$ plotted in Fig.~\ref{fig:03-GatesNoise}(c). Correspondingly, the offline entanglement ratio $R^{(\GP)}_{\rm off}$ [solid red line in Fig.~\ref{fig:03-GatesNoise}(b)] is an $\eta$-independent monotonic function of $T$ that rapidly drops from $1$, whereas $R^{(\GP)}_{\rm on}(\eta)$ [dashed red lines in Fig.~\ref{fig:03-GatesNoise}(b)], albeit being $<1$ for $T=1$, is slower decreasing with the channel losses, so that for $\eta>\eta_{\GP}(T)$ it outperforms case (off).

\section*{Discussion}

\subsection*{Cluster state fusion}\label{sec:Merging}

\begin{figure*}
\includegraphics[width=0.99\linewidth]{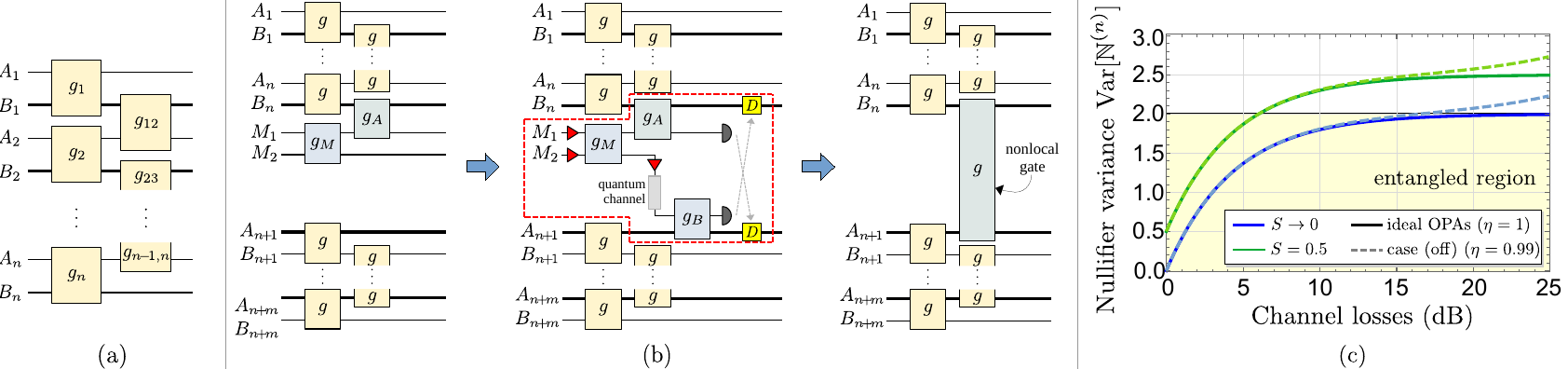}
\caption{(a) QND-type cluster state of $2n$ modes. Ideally, this structure may be generated by the setup in \cite{Yokoyama2013, Larsen2019}, based on multiplexing of optical pulses in time-domain, where (passive) beam splitters are replaced with the (active) QND coupling~(\ref{eq:idealQND}). Within the cluster, we distinguish modes that might describe different bosonic systems, e.g mechanical modes or collective spins, (bold line) from those ones considered optical (normal line). (b) Cluster fusion procedure. We consider two distant cluster states of sizes $2n+2$ between modes $\{A_1,B_1, \ldots, A_n,B_n,M_1,M_2\}$ and $2m$ between $\{ A_{n+1}, B_{n+1}, \ldots,A_{n+m},B_{n+m}\}$, respectively. We then send the edge mode $M_2$ of the first one to the second cluster through an OPA-assisted quantum channel and implement the $\EB$ scheme of Fig.~\ref{fig:01-Protocols}(b). Eventually, we merge the two local clusters into a single bigger distributed cluster of size $2(n+m)$. (c) Nullifier variance ${\rm Var}[\mathbb{N}_{x(p)}^{(n)}]=S+\xi^{(\EB)}$ at the edge node $n$ for different levels of input squeezing $S$ and gain $g=1$. The entangled region is defined by ${\rm Var}[\mathbb{N}_{x(p)}^{(n)}]<2g$. Given the equivalence between $\EB$ and $\SB$ protocols, for ideal OPAs with $\eta=1$, ${\rm Var}[\mathbb{N}_{x(p)}^{(n)}] = S+2g(1-T)$, whereas for channel loss $1-T\ll 1$ and $\eta<1$ the online OPA 3 in Fig.~\ref{fig:01-Protocols}(b) is useless and case (off) provides the optimal strategy, with ${\rm Var}[\mathbb{N}_{x(p)}^{(n)}] \approx S+2g(1-T)-g(1-\eta)/(2\ln(1-T))$ for $1-T,1-\eta\ll 1$, see~(\ref{eq:ExSBeta}).}\label{fig:04-Clusters}
\end{figure*}

The nonlocal quantum interaction schemes above demonstrated can be embedded into CV cluster states, namely multi-mode entangled states providing universal resource for measurement-based quantum computation \cite{Zhang2006, vanLoock2007, Gu2009,Furusawa2019,Budinger2024}. 
QND clusters may be generated by the quantum circuit in Fig.~\ref{fig:04-Clusters}(a), with $2n$ optical modes $\{A_k,B_k\}_k$, $k=1\ldots,n$, coupled in pairs by two sequences of QNDs~(\ref{eq:idealQND}) of gains $g_k$ between $A_k \leftrightarrow B_k$, and then $g_{k,k+1}$ between $B_k \leftrightarrow A_{k+1}$. The resulting output state is characterized by the set of nullifiers, derived in Methods:
\begin{align}\label{eq:NullifierQND}
\begin{cases}
\mathbb{N}_x^{(k)}=x_{A_{k+1}}' - g_{k,k+1} \, x_{B_k}'= x_{A_{k+1}} \,, \\
\mathbb{N}_p^{(k)}=p_{B_k}' + g_{k,k+1} \, p_{A_{k+1}}'= p_{B_k} \,, 
\end{cases}
\end{align}
where $\{x_j,p_j\} (\{x'_j,p'_j\})$ are the input (output) quadratures of modes $j=A_k,B_k$, for which $\mathbb{N}_{x(p)}^{(k)} \to 0$ in the limit of infinite input squeezing, $\langle x_{A_{k+1}}^2\rangle =\langle p_{B_k}^2\rangle=S \to 0$ \cite{Gu2009, Kala2025}.

Unlike beam-splitter-based clusters \cite{Yokoyama2013, Larsen2019}, QND entangled states benefit from quantum erasure \cite{Erasure1,Erasure2,Erasure3,Erasure4}, that allows cluster shaping and shortening by homodynes and displacements \cite{Miwa2010}. Further, here we demonstrate that the nonlocal protocols above presented allow fusion of distinct cluster states \cite{Verstraete2004, Browne2005, Rimock2026}, enabling cluster enlargement at the expense of one (type-I) or two (type-II) qumodes consumption \cite{Xin2023}.
In particular, the topology depicted in Fig.~\ref{fig:04-Clusters}(a) well fits with the $\EB$ protocol [see Fig.~\ref{fig:01-Protocols}(b)], thus implementing type-II fusion. The $\BM$ scheme can be equivalently chosen for the same task, whereas $\SB$ and $\GP$ may be used for type-I fusion of linear cluster states of cascaded QNDs \cite{OBrien2009}.

Let us consider two different QND clusters located at distance, one coupling the $2n+2$ modes $\{A_1,B_1, \ldots, A_n,B_n,M_1,M_2\}$, and the other of size $2m$ between $\{ A_{n+1}, B_{n+1}, \ldots,A_{n+m},B_{n+m}\}$, as in Fig.~\ref{fig:04-Clusters}(b). We fix gains $g_M=-g_0/g_A<0$ and $g_A>0$ for the gates between $M_1 \leftrightarrow M_2$ and $M_1 \leftrightarrow B_n$ [blue boxes in Fig.~\ref{fig:04-Clusters}(b)], respectively, while for the other modes $k=1,\ldots,n+m$ we choose $g_k=g_{k,k+1}=g$ for simplicity. Given this configuration, if we interpret  the last modes of the first cluster as mediators, we recognize the basic topology of the $\EB$ protocol. Therefore, we first prepare $M_{1(2)}$ in two input squeezed states by offline OPAs, then send $M_2$ through an OPA-assisted quantum channel of transmissivity  $T\le 1$ to the second cluster, and finally implement QND of gain $g_B$ with its edge mode $A_{n+1}$, followed by homodyne detection and conditional displacements, so that we effectively merge the two distant clusters into a single bigger distributed cluster of size $2(n+m)$. 

Accordingly, the nullifiers at the edge node $k=n$ become $\mathbb{N}_x^{(n)}= x_{A_{n+1}}' - g \, x_{B_n}'= x_{A_{n+1}}+ {\cal N}_x^{(\EB)}$ and  $\mathbb{N}_p^{(n)}=p_{B_n}' + g \, p_{A_{n+1}}'= p_{B_n} + {\cal N}_p^{(\EB)}$, with the noise operators~(\ref{eq:NoiseEB}). 
A sufficient condition for cluster-type entanglement is then provided by the van Loock-Furusawa criterion \cite{Simon2000, Duan2000, vanLoock2003}:
\begin{align}
{\rm Var}[\mathbb{N}_x^{(k)}] < 2 g \quad \mbox{and} \quad {\rm Var}[\mathbb{N}_p^{(k)}] < 2 g \,,
\end{align}
$k=1,\ldots,n+m$ (see Methods), that for $k\ne n$ implies $S < 2g$, with $S=\langle x_{A_{k+1}}^2\rangle =\langle p_{B_k}^2\rangle$, while at $k=n$ yields:
\begin{align}\label{eq:SimonEX}
{\rm Var}[\mathbb{N}_{x(p)}^{(n)}] = S+ \xi^{(\EB)} < 2g \, , 
\end{align}
$\xi^{(\EB)}=\xi^{(\SB)}$ being the nonlocal-gate excess noise plotted in Fig.~\ref{fig:03-GatesNoise}(a), thus determining the maximum tolerable noise $\xi_{\rm max}= 2g-S$. The nullifier variance at the edge node is plotted in Fig.~\ref{fig:04-Clusters}(c). For ideal lossless OPAs, where $\xi^{(\EB)}=2g(1-T)$, Eq.~(\ref{eq:SimonEX}) conditions the channel transmissivity to $T> S/2g$, preventing long-distance multi-mode entanglement for finite input squeezing $0<S\le1$. 

\subsection*{Conclusions}\label{sec:Concl}
We proposed protocols for a nonlocal (distributed) QND gate between two legitimate users that share a quantum channel, over which a mediator exhibiting quantum features, i.e. squeezing, entanglement, geometric phase effect, is exchanged. The users are also endowed with realistic waveguide OPAs, both employed as squeezers and PSAs for channel-loss compensation. We characterized the gate by logarithmic negativity, proving arbitrary long-distance entanglement for ideal OPAs and robustness to their nonzero internal losses. Finally, we showed application of these gates to merge distinct cluster states into a bigger nonlocal one, thus operating cluster fusion gates at a distance.

We further proved that online OPAs, even if moderately noisy in the realistic settings, are crucial in the large channel-loss regime, being of interest for early-stage experimental demonstrations of the nonlocal gates, to increase the range of maximum tolerable losses and achieve nonzero entanglement. We note that the channel loss can also include imperfect mode matching between the different systems and light. On the other hand, online amplifiers find limited applicability for small channel losses, where non-Gaussian strategies to improve entanglement should be used. In this spirit, our results cover the feasible Gaussian schemes, and therefore provide a relevant benchmark for any non-Gaussian strategy, e.g. non-Gaussian distillation \cite{EntanglementDist1,EntanglementDist2} to be applied to the $\EB$ scheme in Fig.~\ref{fig:01-Protocols}(b) or quantum non-Gaussian error correction \cite{GKPcode, CatCode} in all the considered protocols, if the QND gates are applied on qubit states.
The proposed strategies for distributed gates also hold in the presence of noisy channels, being relevant in the microwave regime. In particular, considering ideal OPAs with $\eta=1$, the resulting excess noise for the $\SB$ and $\GP$ schemes over a thermal-loss quantum channel of transmissivity $T$ and background mean photons $\bar{n}$ become $\xi^{(\SB)}= 2g(1-T)(1+\bar{n})$ and $\xi^{(\GP)}= 2g(1-T)(1+2\bar{n})/(1+T)$, with nonzero entanglement whenever $\xi<2g$. 

The suggested protocols can be naturally embedded into cluster-state technology, extending the CV fusion strategies firstly proposed in \cite{Xin2023} and enabling efficient preparation of large distributed optical networks with different topologies, paving the way for the realization of a global measurement-based CV quantum processor \cite{Danos2007, vanMontfort2025}. Moreover, QND interaction may conveniently replace beam splitters as universal two-mode gate for quantum computation, allowing also for efficient synthesis of nonlocal global nonlinear operations by local non-Gaussian cubic phase gates and nonlocal (Gaussian) QNDs \cite{Budinger2024}. 
The presented entangling methods are also extendable to other platforms for generation of hybrid entanglement between either CV-CV or DV-CV physical systems mediated by optics \cite{Manukhova2020, Thomas2021, Manukhova2022, Manukhova2024,Lei2025, Notarnicola2026}, as well as for quantum transducers, utilizing a mechanical mediator to couple optical modes not directly interacting \cite{Lauk2020, Wang2022,Hou2025,Caleffi2026}.

\section*{Methods}

\subsection*{Derivation of the OPA effective model}\label{app:OPA}
To derive the Heisenberg representation~(\ref{eq:OPAModel}) of realistic OPAs in the presence of distributed losses, we start from the Langevin equations:
\begin{align}
\begin{cases}
\dot{x}= i\, [H_{\rm OPA}, x] - \frac{\gamma}{2} \,x +\sqrt{\gamma} \, x_{\rm in} (t) \,, \nonumber \\[1ex]
\dot{p}= i \,[H_{\rm OPA}, p] - \frac{\gamma}{2} \,p +\sqrt{\gamma} \, p_{\rm in} (t) \, ,
\end{cases}
\end{align}
where $x,p$ are the quadrature operators of the signal mode traveling through the OPA, expressed in shot-noise units, $H_{\rm OPA}= i \chi [ (a\dag)^2-a^2]/2= \chi (xp+px)/4$ is the squeezing Hamiltonian, and $x_{\rm in}(t),p_{\rm in}(t)$ are the time-dependent noisy quadratures of the environment in the the input-output formalism \cite{Serafini2017, Collett1984}. We assume a vacuum bath, such that $\langle x_{\rm in}(t) \rangle= \langle p_{\rm in}(t) \rangle =0$ and $\langle x_{\rm in}(t) x_{\rm in}(t')\rangle= \langle p_{\rm in}(t) p_{\rm in}(t')\rangle= \delta(t-t')$. Then, after evaluating the commutators we find:
\begin{align}
\begin{cases}
\dot{x}= \left(\chi - \frac{\gamma}{2}\right) \,x +\sqrt{\gamma} \, x_{\rm in} (t)  \,, \nonumber \\[1ex]
\dot{p}= - \left(\chi + \frac{\gamma}{2}\right) \,p +\sqrt{\gamma} \, p_{\rm in} (t) \, ,
\end{cases}
\end{align}
with formal solutions \cite{Serafini2017}:
\begin{align}\label{eq:FormalSol}
x(t)&= e^{\left(\chi - \frac{\gamma}{2}\right) t} x(0) + \sqrt{\gamma} \int_0^t dt' e^{\left(\chi - \frac{\gamma}{2}\right) \left(t-t'\right)} x_{\rm in}(t')  \,, \nonumber \\[1ex]
p(t)&= e^{-\left(\chi + \frac{\gamma}{2}\right) t} p(0) + \sqrt{\gamma} \int_0^t dt' e^{-\left(\chi + \frac{\gamma}{2}\right) \left(t-t'\right)} p_{\rm in}(t')  \,,
\end{align}
where $t=L/c$ is the interaction time, fixed by the length $L$ of the waveguide OPA and the light speed $c$ in the medium.
Eq.~(\ref{eq:FormalSol}) yields a Gaussian transformation, therefore it suffices to characterize the output state coming out from the amplifier in terms of its first and second moments. After introducing the notation $x'= x(t), p'= p(t)$ and $x= x(0),p= p(0)$ for the output and input signal modes, and the parameters $G=\exp(2\chi t)$ and $\eta=\exp(-\gamma t)$, we find $\langle x'\rangle= \sqrt{\eta G} \langle x \rangle $, $\langle p'\rangle= \sqrt{\eta/G} \langle p \rangle $, and
\begin{align}
\langle(x')^2 \rangle &= \eta G \langle x^2 \rangle + \gamma  \int_0^t dt' \int_0^t dt'' (\eta G)^{\left(1- \frac{t'+t''}{2t}\right)} \langle x_{\rm in}(t') x_{\rm in}(t'') \rangle \nonumber \\
&=\eta G \langle x^2 \rangle + \gamma  \int_0^t dt' (\eta G)^{\left(1- t'/t \right)} \nonumber \\
&=\eta G \langle x^2 \rangle - \gamma \, t \, \frac{1-\eta G}{\ln (\eta G)} \nonumber \\
&=\eta G \langle x^2 \rangle + (1-\eta) \left[ \frac{1-\eta G}{1-\eta} \frac{\ln(\eta)}{\ln (\eta G)} \right] \, , 
\end{align}
whilst, similarly,
\begin{align}
\langle(p')^2 \rangle =\frac{\eta}{G} \langle p^2 \rangle + (1-\eta) \left[ \frac{1-\eta/G}{1-\eta} \frac{\ln(\eta)}{\ln (\eta/G)} \right]\, .
\end{align}

These results can be interpreted in terms of a mode mixing transformation between a pure-squeezed signal and a (mixed) squeezed thermal state, thus suggesting the effective model in Eq.~(\ref{eq:OPAModel}): a sequence of an ideal OPA, performing pure squeezing of gain $G>0$, coupled by a beam splitter of transmissivity $\eta\le 1$ to a squeezed-thermal bath state having $\langle x_{\nn} \rangle = \langle p_{\nn} \rangle=0$ and the $\langle x_{\nn}^2 \rangle$ and $\langle p_{\nn}^2 \rangle$ of Eq.~(\ref{eq:xnnpnn}).

\subsection*{Extended calculations for the nonlocal QND schemes}\label{app:NONLOCALAPP}

\begin{figure*}
\includegraphics[width=0.99\linewidth]{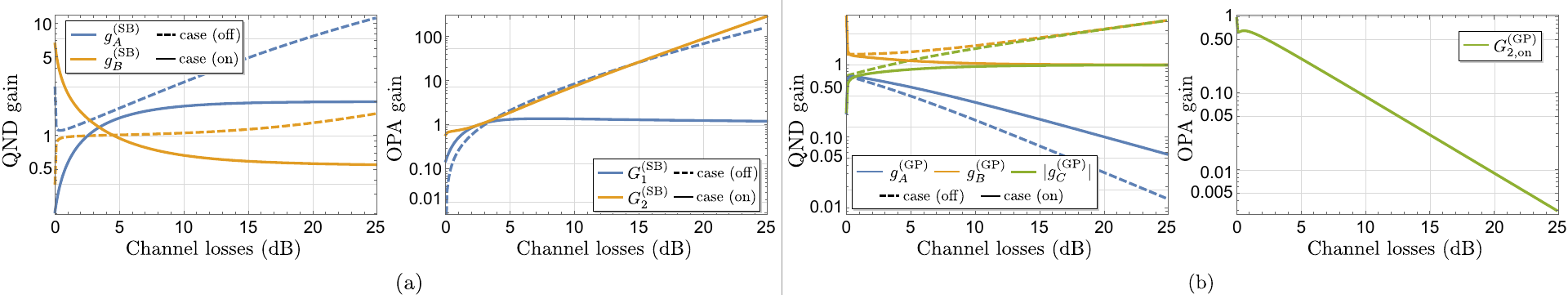}
\caption{Log plot of the optimized parameters for the $\SB$ (a) and $\GP$ (b) protocols for target gain $g=1$ as a function of the channel losses in cases (off) and (on), respectively, and $\eta=0.9$. In the $\GP$ scheme, the optimal configuration of OPA 1 is obtained in the limit $G_{1,{\rm k}}^{(\GP)} \to 0$, $\rm k= off,on$, corresponding to large position-like squeezing of the mediator $M$.}\label{fig:05-OptGains}
\end{figure*}

We present the extended derivation of the modes linear transformation~(\ref{eq:NonLocalTrans}) associated with the discussed nonlocal QND protocols.
We start with the $\SB$ scheme in Fig.~\ref{fig:01-Protocols}(a). First, the mediator $M$ is squeezed by OPA 1 and coupled to $A$ by QND of gain $g_A>0$, leading to:
\begin{align}
x_A^{\rm (i)}&= x_A \,, \quad   p_A^{\rm (i)}= p_A - g_A \left(\sqrt{\frac{\eta_1}{G_1}} p_M + \sqrt{1-\eta_1} p_{\nn ,1}\right) \, , \nonumber  \\[1ex]
x_M^{\rm (i)}&= \sqrt{\eta_1 G_1} x_M + \sqrt{1-\eta_1} x_{\nn ,1}+ g_A x_A \,, \quad p_M^{\rm (i)}= p_M \, .
\end{align}
Thereafter, Alice transmits $M$ to Bob through the channel, using OPA 2 for pre-amplification, and Bob performs a second QND of gain $g_B= g/(g_A \sqrt{\eta_2 G_2 T})$, obtaining:
\begin{align}
x_M^{\rm (ii)}&= \sqrt{\eta_2 G_2 T} x_M^{\rm (i)} + \sqrt{(1-\eta_2) T} x_{\nn ,2} + \sqrt{1-T} x_\ch \, , \nonumber  \\[1ex]
p_M^{\rm (ii)}&= \sqrt{\frac{\eta_2 T}{G_2}} p_M + \sqrt{(1-\eta_2) T} p_{\nn ,2} + \sqrt{1-T} p_\ch - g_B p_B \,, \nonumber  \\[1ex]
x_B^{\rm (ii)}&=x_B + g_B x_M^{\rm (ii)} \,, \qquad p_B^{\rm (ii)}= p_B \, .
\end{align}
Finally, Bob measures $p_M^{\rm (ii)}$ and communicate the measurement outcome $\bar{p}$ to Alice, who applies a momentum displacement of amplitude $g_A \sqrt{\eta_2 G_2 T} \bar{p}$, that on average implements the Heisenberg transformation $x_A^{\rm (ii)}=x_A^{\rm (i)}$ and $p_A^{\rm (ii)} = p_A^{\rm (i)} + g_A \sqrt{\eta_2 G_2 T} p_M^{\rm (ii)}$. Explicit substitution yields Eq.~(\ref{eq:NoiseSB}). 

With similar method, we approach the $\EB$ protocol of Fig.~\ref{fig:01-Protocols}(b). We prepare the entangled resource state between $M_1$ and $M_2$, by first squeezing the mediators:
\begin{align}
x_{M_k}^{\rm (i)}&= \sqrt{\eta_k G_k} x_{M_k} + \sqrt{1-\eta_k} x_{\nn ,k} \,, \nonumber \\ 
p_{M_k}^{\rm (i)}&= \sqrt{\frac{\eta_k}{G_k}} p_{M_k} + \sqrt{1-\eta_k} p_{\nn ,k} \,, \quad k=,1,2\, ,
\end{align}
and then performing QND of gain $g_M=-g_0/g_A<0$:
\begin{align}
x_{M_1}^{\rm (ii)}&= x_{M_1}^{\rm (i)} \,, \qquad p_{M_1}^{\rm (ii)}= p_{M_1}^{\rm (i)} +\frac{g_0}{g_A} p_{M_2}^{\rm (i)} \,, \nonumber \\ 
x_{M_2}^{\rm (ii)}&= x_{M_2}^{\rm (i)} - \frac{g_0}{g_A} x_{M_1}^{\rm (i)}  \,, \qquad p_{M_2}^{\rm (ii)}= p_{M_2}^{\rm (i)} \, .
\end{align}
Subsequently, $M_2$ is sent through the channel, with OPA 3 as PSA, and Alice and Bob implement QNDs of gain $g_A>0$ and $g_B= g/(g_A \sqrt{\eta_3 G_3 T})$, so that:
\begin{align}
x_A^{\rm (iii)}&= x_A \,, \qquad p_A^{\rm (iii)}= p_A - g_A p_{M_1}^{\rm (ii)} \, , \nonumber \\ 
x_{M_1}^{\rm (iii)}&=x_{M_1}^{\rm (i)} + g_A x_A \, , \qquad p_{M_1}^{\rm (iii)}= p_{M_1}^{\rm (ii)}
\end{align}
and
\begin{align}
x_{M_2}^{\rm (iii)}&= \sqrt{\eta_3 G_3 T} x_{M_2}^{\rm (ii)} + \sqrt{(1-\eta_3)T} x_{\nn,3} + \sqrt{1-T} x_\ch \,, \nonumber \\
p_{M_2}^{\rm (iii)}&= \sqrt{\frac{\eta_3 T}{G_3}} p_{M_2}^{\rm (i)} + \sqrt{(1-\eta_3)T} p_{\nn,3} + \sqrt{1-T} p_\ch - g_B p_B \, , \nonumber \\
x_{B}^{\rm (iii)}&= x_B + g_B x_{M_2}^{\rm (iii)}  \,, \qquad p_B^{\rm (iii)}= p_B \, .
\end{align}
Finally, we measure $x_{M_1}^{\rm (iii)}$ and $p_{M_2}^{\rm (iii)}$, and displace $p_A^{\rm (iii)}$ and $x_B^{\rm (iii)}$ to get $p_A^{\rm (iv)}= p_A^{\rm (iii)} + g_0 \sqrt{\eta_3 G_3 T} p_{M_2}^{\rm (iii)} $ and $x_B^{\rm (iv)}= x_B^{\rm (iii)} + g x_{M_1}^{\rm (iii)}/g_A$, leading to~(\ref{eq:NoiseEB}). 

We now move to the $\BM$ setup of Fig.~\ref{fig:01-Protocols}(c). Like the $\EB$, we first offline squeeze mediators $M_{1(2)}$ by OPAs 1(2):
\begin{align}
x_{M_k}^{\rm (i)}&= \sqrt{\eta_k G_k} x_{M_k} + \sqrt{1-\eta_k} x_{\nn ,k} \,, \nonumber \\ 
p_{M_k}^{\rm (i)}&= \sqrt{\frac{\eta_k}{G_k}} p_{M_k} + \sqrt{1-\eta_k} p_{\nn ,k} \,, \quad k=,1,2\, .
\end{align}
Thereafter, Alice and Bob performs local QNDs with $M_{1(2)}$, respectively, such that:
\begin{align}
x_A^{\rm (ii)}&= x_A \,, \qquad p_A^{\rm (ii)}= p_A- g_A p_{M_1}^{\rm (i)} \,, \nonumber \\ 
x_{M_1}^{\rm (ii)}&= x_{M_1}^{\rm (i)} + g_A x_A  \,, \qquad p_{M_1}^{\rm (ii)}= p_{M_1}^{\rm (i)} \,, \nonumber \\[2ex]
x_{M_2}^{\rm (ii)}&= x_{M_2}^{\rm (i)} \,, \qquad p_{M_2}^{\rm (ii)}= p_{M_2}^{\rm (i)} - g_B p_B \,, \nonumber \\ 
x_B^{\rm (ii)}&= x_B + g_B x_{M_2}^{\rm (i)}  \,, \qquad p_B^{\rm (ii)}= p_B \, .
\end{align}
Mediator $M_1$ is then sent to Bob by the OPA-3 assisted channel, namely:
\begin{align}
x_{M_1}^{\rm (iii)}&= \sqrt{\eta_3 G_3 T} x_{M_1}^{\rm (ii)} + \sqrt{(1-\eta_3)T} x_{\nn ,3} + \sqrt{1-T} x_\ch \,,\nonumber \\
p_{M_1}^{\rm (iii)}&= \sqrt{\frac{\eta_3  T}{G_3}} p_{M_1}^{\rm (ii)}  + \sqrt{(1-\eta_3)T} p_{\nn ,3} + \sqrt{1-T} p_\ch  \,, 
\end{align}
and undergoes the Bell measurement with $M_2$, consisting of a balanced beam splitter interaction such that:
\begin{align}
x_{M_1}^{\rm (iv)}&= \frac{x_{M_1}^{\rm (iii)} + x_{M_2}^{\rm (iii)}}{\sqrt{2}}  \, , \qquad p_{M_1}^{\rm (iv)}= \frac{p_{M_1}^{\rm (iii)} + p_{M_2}^{\rm (iii)}}{\sqrt{2}}\,,\nonumber \\
x_{M_2}^{\rm (iv)}&= \frac{x_{M_2}^{\rm (iii)} - x_{M_1}^{\rm (iii)}}{\sqrt{2}}  \, , \qquad p_{M_2}^{\rm (iv)}= \frac{p_{M_2}^{\rm (iii)} - p_{M_1}^{\rm (iii)}}{\sqrt{2}}\,,
\end{align}
followed by homodynes of $p_{M_1}^{\rm (iv)}$ and $x_{M_2}^{\rm (iv)}$.  The last operations are the conditonal displacements $p_A^{\rm (v)}=p_A^{\rm (ii)} +g_A \sqrt{2 \eta_3 G_3 T} p_{M_1}^{\rm (iv)}$ and $x_B^{\rm (v)}= x_B^{\rm (ii)} -\sqrt{2} g_B \, x_{M_2}^{\rm (iv)}$, that give~(\ref{eq:NoiseBM}).
 
Instead, in the $\GP$ scheme [Fig.~\ref{fig:01-Protocols}(d)], we begin by squeezing $M$ by OPA 1 and performing QND of gain $g_A>0$:
\begin{align}
x_A^{\rm (i)}&= x_A \,, \quad   p_A^{\rm (i)}= p_A - g_A \left(\sqrt{\frac{\eta_1}{G_1}} p_M + \sqrt{1-\eta_1} p_{\nn ,1}\right) \, , \nonumber  \\[1ex]
x_M^{\rm (i)}&= \sqrt{\eta_1 G_1} x_M + \sqrt{1-\eta_1} x_{\nn ,1}+ g_A x_A \,, \quad p_M^{\rm (i)}= p_M \, ,
\end{align}
then let the mediator mode pass through the channel for the first time and implement QND with $B$ of gain $g_B>0$:
\begin{align}
x_M^{\rm (ii)}&= \sqrt{\eta_2 G_2 T} x_M^{\rm (i)} + \sqrt{(1-\eta_2) T} x_{\nn ,2} + \sqrt{1-T} x_{\ch,1} \, , \nonumber  \\[1ex]
p_M^{\rm (ii)}&= \sqrt{\frac{\eta_2 T}{G_2}} p_M + \sqrt{(1-\eta_2) T} p_{\nn ,2} + \sqrt{1-T} p_{\ch,1} - g_B p_B \,, \nonumber  \\[1ex]
x_B^{\rm (ii)}&=x_B + g_B x_M^{\rm (ii)} \,, \qquad p_B^{\rm (ii)}= p_B \, ,
\end{align}
and finally transmit it back to Alice by the second passage through the channel:
\begin{align}
x_M^{\rm (iii)}&= \sqrt{\eta_3 G_3 T} x_M^{\rm (ii)} + \sqrt{(1-\eta_3) T} x_{\nn ,3} + \sqrt{1-T} x_{\ch,2} \, , \nonumber  \\[1ex]
p_M^{\rm (iii)}&= \sqrt{\frac{\eta_3 T}{G_3}} p_M^{\rm (ii)} + \sqrt{(1-\eta_3) T} p_{\nn ,3} + \sqrt{1-T} p_{\ch,2}  \,. 
\end{align}
Now, $A$ is coupled to $M$ by the last QND of gain $g_C=-g \sqrt{G_3/(\eta_3 T g_B^2)}<0$:
\begin{align}
x_A^{\rm (iv)}&=x_A  \,, \qquad p_A^{\rm (iv)}= p_A^{\rm (i)} - g_C p_M^{\rm (iii)} \, , \nonumber  \\[1ex]
x_M^{\rm (iv)}&= x_M^{\rm (iii)} + g_C \, x_A  \,, \qquad p_M^{\rm (iv)}= p_M^{\rm (iii)} \, ,
\end{align}
then we measure $x_M^{\rm (iv)}$ and displace $x_B^{\rm (ii)}$ by $x_B^{\rm (iv)}=x_B^{\rm (ii)}+g_B \Gamma x_M^{\rm (iv)}$, retrieving~(\ref{eq:NoiseGP}). 

In all four cases, we effectively implement the additive-Gaussian-noise map ${\bf r}'= \mathbb{S}\, {\bf r} + {\bf N}^{(\rm s)}$, ${\bf r}=(x_A,p_A,x_B,p_B)$, where 
\begin{align}
\mathbb{S}= \begin{pmatrix}
1 & 0 & 0 & 0 \\
0 &1 & 0 & -g \\
g & 0 & 1 & 0 \\
0 & 0 & 0 &  1
\end{pmatrix}
\end{align}
is the symplectic matrix associated with ideal QND~(\ref{eq:idealQND}) and ${\bf N}^{(\rm s)}= (0,{\cal N}_p^{(\rm s)}, {\cal N}_x^{(\rm s)},0)$, ${\rm s}=\SB,\EB,\BM,\GP$.
As a consequence, if system $AB$ is initially prepared in a Gaussian state with covariance matrix (CM) $\bmsigma_{\rm in}= (\langle {\bf r}_j {\bf r}_k + {\bf r}_k {\bf r}_j \rangle/2)_{jk}$, the output bipartite state is still Gaussian with CM:
\begin{align}
\bmsigma_{\rm out}= \mathbb{S} \bmsigma_{\rm in} \mathbb{S}^{\sf T} + \bmsigma_{\nn}^{(\rm s)} \, ,
\end{align}
where:
\begin{align}
\bmsigma_{\nn}^{(\rm s)}= \begin{pmatrix}
0 & 0 & 0 & 0 \\
0 & \xi^{(\rm s)}_p & 0 & 0 \\
0 & 0 & \xi^{(\rm s)}_x & 0 \\
0 & 0 & 0 &  0
\end{pmatrix}
\end{align}
is the excess noise CM associated with the additional modes, and $\xi^{(\rm s)}_{p(x)}= {\rm Var}[{\cal N}_{p(x)}^{(\rm s)}]$. In particular, for input vacuum states considered in the main text, we have $\bmsigma_{\rm in}= \mathbbm{1}_4$ and:
\begin{align}
\bmsigma_{\rm out}
&= \begin{pmatrix}
 \bmsigma_A & \bmsigma_{AB} \\
  \bmsigma_{AB}^{\sf T} &  \bmsigma_{B} \\
\end{pmatrix}= 
\begin{pmatrix}
 1 & 0 & g & 0 \\
 0 & 1+g^2+\xi^{(\rm s)}_p & 0 & -g \\
 g & 0 & 1+g^2+\xi^{(\rm s)}_x & 0 \\
 0 & -g & 0 & 1 \\
\end{pmatrix} \, .
\end{align}

Logarithmic negativity is then computed as $E_N^{(\rm s)}= \max\{0,-\ln( \tilde{d}_-)\}$, where \cite{Ferraro2005,Olivares2012}:
\begin{align}
\tilde{d}_-=\sqrt{\frac{\tilde{\Delta} -\sqrt{\tilde{\Delta}^2-4  \det(\bmsigma_{\rm out})}}{2}}
\end{align}
is the smallest partially-transposed symplectic eigenvalue, with $\tilde{\Delta}= I_A+I_B-2I_{AB}$, $I_{k}=\det(\bmsigma_{k})$, $k=A,B,AB$, and is maximized in the symmetric excess noise configuration. In turn, for all the discussed protocols, the local QND and OPA gains are numerically optimized to maximize entanglement (or, equivalently, minimize the excess noise) with the additional constraint $\xi^{(\rm s)}_p=\xi^{(\rm s)}_x=\xi^{(\rm s)}$, leading to~(\ref{eq:LN}).
The optimized parameters for both the $\SB$ and $\GP$ schemes are discussed in the main text in the case of ideal OPAs, whereas for realistic amplifiers with $\eta <1$ we plot them in Fig.~\ref{fig:05-OptGains}. The $\EB$ protocol is equivalent to $\SB$ in the limit $g_A^2 [\eta_1/G_1 + (1-\eta_1/G_1) \ln(\eta_1)/\ln(\eta_1/G_1)] \ll 1$, while the $\BM$ equals $\SB$ when $[\eta_2/G_2 +(1-\eta_2/G_2) \ln(\eta_2)/\ln(\eta_2/G_2) \rangle ]/g_B^2 \ll 1$.

\subsection*{Impact of non-unit incoupling OPA losses}
\begin{figure*}
\includegraphics[width=0.99\linewidth]{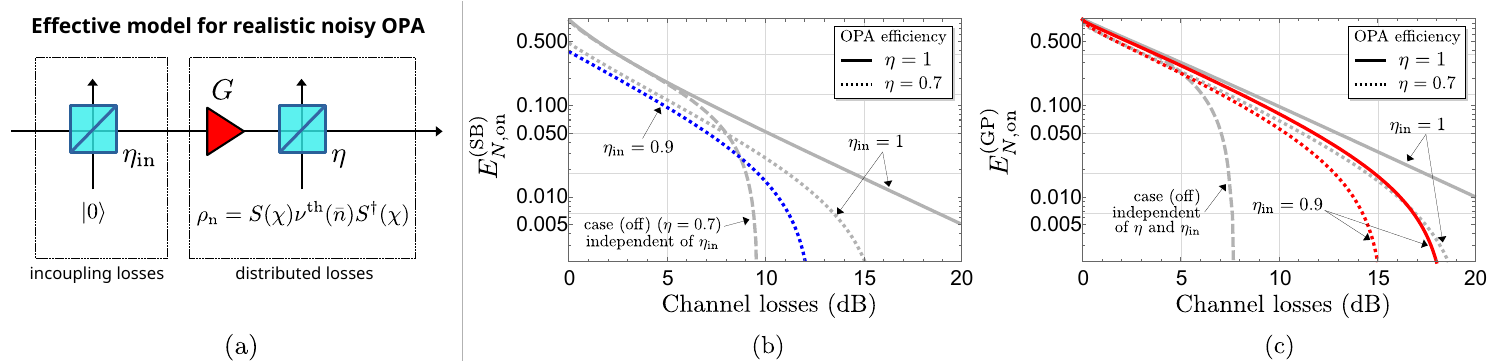}
\caption{(a) Schematic model of realistic noisy OPA, including incoupling losses, described by pure-loss channel with transmissivity $\eta_{\rm in}\le 1$ and distributed losses, modeled in terms of pure squeezing followed by squeezed-thermal-loss channel of transmissivity $\eta\le 1$ and a suitable squeezed-thermal bath $\rho_\nn$, as derived in App.~\ref{app:OPA}. (b-c) Logarithmic negativity $E_{N,{\rm on}}^{(\rm s)}$, ${\rm s}=\SB,\GP$, in the presence of incoupling losses of different values of incoupling efficiency $\eta_{\rm in}$ and distributed loss $\eta$, respectively. When $\eta=1$, the $\SB$ protocol can be made independent of $\eta_{\rm in}$, provided to use only offline OPA~1 with optimized gain, whereas the $\GP$ scheme crucially needs online OPA~2, that for $\eta_{\rm in}<1$ makes entanglement vanish at large transmission losses.}\label{fig:06-InLoss}
\end{figure*}

Beyond distributed losses inside the OPA, that are expected to be the prevalent noise source for large squeezing generation, another crucial limiting factor in view of practical realizations of waveguide OPAs is provided by the incoupling losses, being consequence of imperfect signal injection into the $\chi^{(2)}$ waveguide \cite{Incoupling}. By combining these two effects, a realistic noisy OPA turns out to be described by the effective model in Fig.~\ref{fig:06-InLoss}(a), that is a pure-loss channel of transmissivity $\eta_{\rm in}\le 1$ modeling incoupling losses, followed by the squeezed-thermal-loss channel for the internal distributed losses derived in App.~\ref{app:OPA}.

By including this further element in the nonlocal QND protocols of Fig.~\ref{fig:01-Protocols}, the noise modes in Eq.~(\ref{eq:NonLocalTrans}) now become:
\begin{subequations}
\begin{align}
&{\cal N}_p^{(\SB)}= - g_A  (1- \eta_2 \eta_{\rm in,2} T) \Bigg( \sqrt{\frac{\eta_1 \eta_{\rm in,1}}{G_1}} p_M +\sqrt{\frac{\eta_1 (1-\eta_{\rm in,1})}{G_1}} p_{\rm in,1}  \nonumber \\
&\hspace{.5cm}+  \sqrt{1-\eta_1} p_{\nn ,1} \Bigg) + g_A \eta_2 T \sqrt{\eta_{\rm in,2} (1-\eta_{\rm in,2})} p_{\rm in,2} \nonumber \\
&\hspace{.5cm}+ g_A \sqrt{  \eta_2 \eta_{\rm in,2} G_2 T} \left( \sqrt{1-T} p_\ch + \sqrt{(1-\eta_2)T} p_{\nn,2} \right)  
\end{align}
\begin{align}
&{\cal N}_x^{(\SB)}=\frac{g}{g_A}  \Big( \sqrt{\eta_1\eta_{\rm in,1}  G_1} x_M +  \sqrt{\eta_1(1-\eta_{\rm in,1})  G_1} x_{\rm in,1} \nonumber \\
&\hspace{.5cm}+ \sqrt{1-\eta_1} x_{\nn ,1} \Big) + \frac{g}{g_A} \sqrt{ \frac{1-\eta_{\rm in,2}}{\eta_{\rm in,2}} } x_{\rm in,2} \nonumber \\
&\hspace{.5cm}+ \frac{g}{g_A \sqrt{ \eta_2 \eta_{\rm in,2} G_2 T}} \left( \sqrt{1-T} x_\ch + \sqrt{(1-\eta_2)T} x_{\nn,2} \right) \, ,
\end{align}
\end{subequations}
for the $\SB$ protocol, while for the $\GP$ scheme with the mediator-independence conditions~(\ref{eq:ConditionGP}):
\begin{subequations}\label{eq:NoiseGP}
\begin{align}
{\cal N}_p^{(\GP)}&= \frac{g \sqrt{1-T}}{g_B} \left( p_{\ch,1} + \sqrt{\frac{G_2}{\eta_2 \eta_{\rm in,2}T}} p_{\ch,2}\right)  \nonumber \\
&\hspace{.5cm}+ \frac{g}{g_B} \left(\sqrt{\frac{G_2(1-\eta_2)}{\eta_2\eta_{\rm in,2}}} p_{\nn,2} + \sqrt{\frac{1-\eta_{\rm in,2}}{\eta_{\rm in,2}}} p_{\rm in,2} \right)\, ,
\end{align}
\begin{align}
{\cal N}_x^{(\GP)}&= g_B \left(1+ \Gamma \sqrt{\eta_2\eta_{\rm in,2} G_2 T} \right)  \Big[ \sqrt{(1-\eta_1)T} x_{\nn ,1}   \nonumber \\
&+ \sqrt{1-T} x_{\ch,1} \Big]  + g_B \Gamma \Big( \sqrt{1-T} x_{\ch,2} +  \sqrt{(1-\eta_2)T} x_{\nn,2} \nonumber \\
&+ \sqrt{\eta_2 (1-\eta_{\rm in,2}) G_2 T} x_{\rm in,2} \Big) \, ,
\end{align}
\end{subequations}
where quadratures $x_{\rm in},p_{\rm in}$ describe the additional mode associated with the incoupling losses, and
$$\Gamma= \sqrt{\frac{\eta_2 \eta_{\rm in,2}  T}{G_2} }\frac{1-T}{\eta_2 \eta_{\rm in,2}T^{5/2} -1}\, .$$
The equivalence between the $\EB (\BM)$ and the $\SB$ schemes is still preserved. 

We also note that, for both the $\SB$ and the optimal $\GP$ protocols, the performance of offline OPA~1 is independent of the incoupling efficiency, as in the former case we have a vacuum mediator, while in the latter the large position-like squeezing $G_1\to 0$ makes the scheme independent of the initial mediator state. In turn, the results presented in the main text for case (off), i.e. where only OPA~1 is considered, are not affected if $\eta_{\rm in}=\eta_{\rm in,1}<1$.
Instead, incoupling efficiency becomes relevant for case (on), when the noisy online OPA~2 is maintained. In this case, incoupling losses induce a reduction of logarithmic negativity, see Fig.~\ref{fig:06-InLoss}(b-c), with different impact for $\SB$ and $\GP$. In the $\SB$ protocol, the entanglement reduction is observed only if the internal OPA efficiency is $\eta<1$, with a consequently lower value of maximum transmission loss. However, if $\eta=1$, given the optimal condition $G_1 G_2=(1-T)/T$, online OPA~2 can be removed if the OPA~1 gain $G_1$ is suitably tuned to $G_1=(1-T)/T$, thus retrieving the $\eta_{\rm in}$-independent results of the ideal case, that allow arbitrary long-distance entanglement. On the other hand, in the $\GP$ scheme, incoupling losses of online OPA~2 make $E_{N,{\rm on}}^{(\rm GP)}$ drop to 0 at large transmission losses even if the internal efficiency is $\eta=1$. Now, OPA~2, being crucial for loss-compensation of the second channel, unavoidably amplifies the additional incoupling loss mode $p_{\rm in,2}$, inducing the entanglement break. 

\subsection*{Impact of finite OPA gain in the GP protocol}

\begin{figure}
\includegraphics[width=0.95\columnwidth]{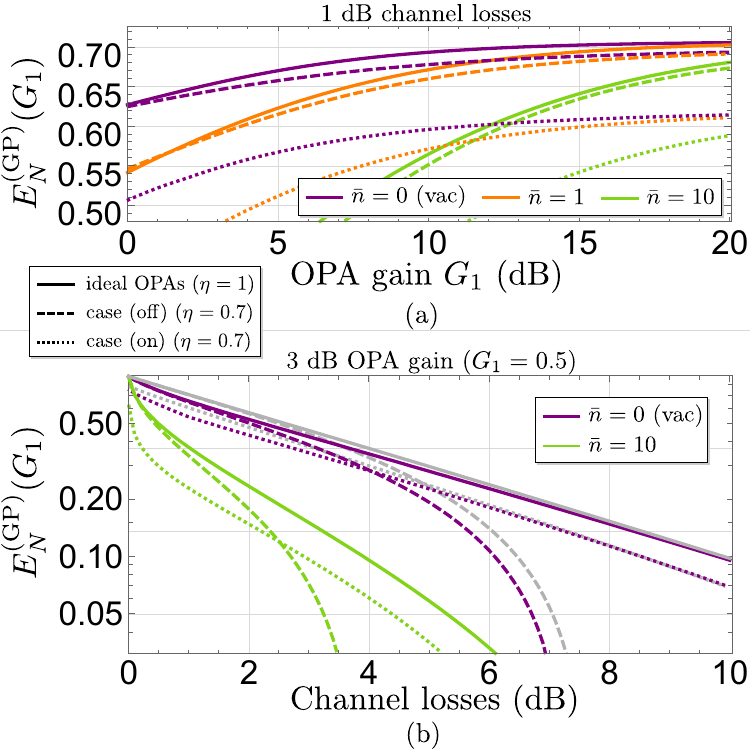}
\caption{(a) Logaritmic negativity $E_N^{(\GP)}(G_1)$ for target gain $g=1$ and 1 dB channel losses as a function of the OPA~1 gain $G_1$ for different mean thermal photons $\bar{n}$ of the mediator. For large offline squeezing $G_1\ll 1$, the curves converge to the protocol described in the main text and depicted in Fig.~\ref{fig:02-GatesNeg}. (b) Logaritmic negativity $E_N^{(\GP)}(G_1)$ for target gain $g=1$ and fixed OPA~1 gain $G_1=0.5$ (i.e. 3 dB), as a function of the channel losses. The light-gray lines are associated with the protocol of main text and Fig.~\ref{fig:02-GatesNeg}. For vacuum mediator, large OPA~1 gain is needed only for $1-T\ll1$, while for large losses $E_N^{(\GP)}(G_1)$ approaches the target protocol. On the contrary, with thermal mediator large offline squeezing is crucial in all regimes to both increase negativity for small $T$ and prevent entanglement break at large $T$. In both the figures, solid lines refer to lossless OPAs with unit efficiency, while dashed and dotted ones indicate the performance of realistic OPAs ($\eta=0.7$) for the two cases of only offline squeezing and both online and offline OPAs, respectively.}\label{fig:07-FiniteG}
\end{figure}

As discussed in the main text, the $\GP$ protocol yields its maximum performance in the limit of large offline squeezing, corresponding to the OPA~1 gain $G_1\to 0$, see Fig.~\ref{fig:01-Protocols}(d). If so, the protocol becomes independent of the mediator, that could be prepared in any arbitrary quantum state $\rho_M$. Here, we address robustness of $\GP$ to the more realistic scenario of limited offline amplification, namely in the presence of fixed (nonzero) gain $G_1>0$, where, now, the mediator-independence conditions~(\ref{eq:ConditionGP}) do not hold and we expect dependence of the scheme on state $\rho_M$.
To this aim, we consider the mediator in a thermal state $\rho_M=\sum_{n=0}^{\infty} \bar{n}^n/(\bar{n}+1)^{n+1} |n\rangle\langle n|$ of $\bar{n}\ge 0$ average photons, being a typical configuration observed in quantum transducers. Case $\bar{n}=0$ then yields the vacuum state  $\rho_M=|0\rangle\langle 0|$.
To qualify the gate, we compute the logarithmic negativity $E_N^{(\GP)}(G_1)$ as a function of both gain $G_1$ and channel transmissivity $T$, by maximizing Eq.~(\ref{eq:LN}) over the free parameters $(g_A,g_B,G_2)$, with the excess noise~(\ref{eq:NoiseGP}). As in the main text, we address both the cases of ideal lossless OPAs and lossy amplifiers with efficiency $\eta<1$ in cases (off) and (on), corresponding to the use of only offline/both online and offline squeezers, respectively. For the sake of simplicity, we assume perfect incoupling procedure.

$E_N^{(\GP)}(G_1)$ is a monotonic function of $G_1$, see Fig.~\ref{fig:07-FiniteG}(a), with asymptotic trend for large enough amplification gain, so that $E_N^{(\GP)}(G_1) \approx E_N^{(\GP)}$ for $G_1\to0$, with the $E_N^{(\GP)}$ of Fig.~\ref{fig:02-GatesNeg}. Convergence to this regime is faster for a vacuum mediator, while a larger thermal occupancy $\bar{n} >0$ requires larger amount of offline squeezing to be compensated.
In turn, we observe different behaviors of the logarithmic negativity as a function of the channel losses for fixed gain $G_1$, as plotted in Fig.~\ref{fig:07-FiniteG}(b). With vacuum mediator and ideal OPAs ($\eta=1$), $E_N^{(\GP)}(G_1)$ is smaller than $E_N^{(\GP)}$ only when $1-T\ll 1$, whilst for $T\ll 1$, $E_N^{(\GP)}(G_1) \approx E_N^{(\GP)}$, proving large offline squeezing to be crucial only in the small loss limit. Analogous considerations hold for lossy OPAs, with $E_{N,{\rm off}}^{(\GP)}(G_1)<E_{N,{\rm off}}^{(\GP)}$, while $E_{N,{\rm on}}^{(\GP)}(G_1) \approx E_{N,{\rm on}}^{(\GP)}$ for small $T$.
On the other hand, in the presence of a thermal mediator the logarithmic negativity deviates from the infinite-offline-squeezing protocol for all $T$, and for finite squeezing $G_1>0$ it drops to $0$ even with lossless OPAs, thus exhibiting a maximum transmission loss that occurs whenever the corresponding excess noise beats the maximum level $\xi_{\rm max}=2g$. Therefore, in this latter case, a large amplification gain is needed to efficiently distribute entanglement at distance.

\subsection*{Nullifiers for QND clusters}\label{app:Nullifiers}

In this Appendix, we characterize the QND-type cluster state schematized in Fig.~\ref{fig:04-Clusters}(a). We first derive the overall quadrature transformation in the Heisenberg picture. Initially, the input modes $\{A_k,B_k\}_k$, $k=1,\ldots, n$, undergo a first round of QNDs of gain $g_k$ between $A_k$ and $B_k$:
 \begin{align}
x_{A_k}^{\rm (i)}&= x_{A_k} \,, \qquad   p_{A_k}^{\rm (i)}= p_{A_k} - g_k \, p_{B_k} \, , \nonumber  \\[1ex]
x_{B_k}^{\rm (i)}&= x_{B_k} + g_k \, x_{A_k} \,, \qquad p_{B_k} ^{\rm (i)}= p_{B_k}  \, .
\end{align} 
Then, mode $B_k$ is further coupled to $A_{k+1}$ by QND of gain $g_{k,k+1}$:
\begin{align}
x_{B_k}^{\rm (ii)}&= x_{B_k}^{\rm (i)} \,, \qquad p_{B_k} ^{\rm (ii)}= p_{B_k}^{\rm (i)} - g_{k,k+1} p_{A_{k+1}}^{\rm (i)}  \, , \nonumber  \\[1ex]
x_{A_{k+1}}^{\rm (ii)}&= x_{A_{k+1}}^{\rm (i)} +g_{k,k+1} x_{B_k}^{\rm (i)}   \,, \qquad p_{A_{k+1}} ^{\rm (ii)}= p_{A_{k+1}}^{\rm (i)} \,,
\end{align} 
leading to the overall transformation:
\begin{align}\label{eq:HeisenbergClusters}
\begin{cases}
x_{A_k}'=x_{A_k}+ g_{k-1,k} \left(x_{B_{k-1}} + g_{k-1} \, x_{A_{k-1}} \right) \,,  \\[.15ex]
p_{A_k}'=p_{A_k} - g_k \, p_{B_k} \,,  \\[.15ex]
x_{B_k}'=x_{B_k} + g_k \, x_{A_k}  \, , \\[.15ex]
p_{B_k}'=p_{B_k}- g_{k,k+1} \left(p_{A_{k+1}} - g_{k+1} \, p_{B_{k+1}} \right) \,.
\end{cases}
\end{align} 

The corresponding nullifiers $\mathbb{N}_{x(p)}^{(k)} $ are identified by inverting the linear transformation~(\ref{eq:HeisenbergClusters}), leading to Eq.~(\ref{eq:NullifierQND}), so that the cluster state is the zero-eigenvalue eigenstate of  $\mathbb{N}_{x(p)}^{(k)} $ in the limit $\langle x_{A_{k+1}}^2\rangle =\langle p_{B_k}^2\rangle\to 0$ for all $k$, corresponding to infinite input squeezing. In more realistic conditions, a sufficient condition for multi-mode entanglement is derived following the van Loock-Furusawa approach \cite{vanLoock2003} that extends Simon criterion \cite{Simon2000}, leading to:
\begin{align}
{\rm Var} [\mathbb{N}_{x}^{(k)}]  + {\rm Var} [\mathbb{N}_{p}^{(k)}] < 4 |g_{k,k+1}|  \,,
\end{align} 
expressed in shot-noise units, that, assuming equal variances ${\rm Var} [\mathbb{N}_{x}^{(k)}]={\rm Var} [\mathbb{N}_{p}^{(k)}]$ gives the two independent conditions in Eq.~(\ref{eq:SimonEX}).

Finally, in the cluster state fusion procedure of Fig.~\ref{fig:04-Clusters}(b), we start with two independent clusters described by transformations~(\ref{eq:HeisenbergClusters}) and then implement the $\EB$ protocol outlined in App.~\ref{app:NONLOCALAPP}. Eventually, we find a single cluster of size $2(n+m)$, being still described by transformation~(\ref{eq:HeisenbergClusters}) for $k\ne n, n+1$, while at the merged nodes $k=n, n+1$ we now have:
\begin{align}
& x_{A_{n+1}}'=x_{A_{n+1}}+ g_{n,n+1} \left(x_{B_{n}} + g_{n} \, x_{A_{n}} \right) + {\cal N}_x^{(\EB)} \,,\nonumber  \\
& p_{A_{n+1}}'=p_{A_{n+1}} - g_{n+1} \, p_{B_{n+1}} \,, \nonumber \\
& x_{B_{n}}'=x_{B_{n}} + g_{n} \, x_{A_n}  \, ,\nonumber \\
& p_{B_{n}}'=p_{B_{n}}- g_{n,n+1} \left(p_{A_{n+1}} - g_{n+1} \, p_{B_{n+1}}  \right) + {\cal N}_p^{(\EB)}\,,
\end{align}
with the noise operators~(\ref{eq:NoiseEB}), so that the corresponding nullifiers become $\mathbb{N}_x^{(n)}= x_{A_{n+1}}' - g_{n,n+1} \, x_{B_n}'= x_{A_{n+1}}+ {\cal N}_x^{(\EB)}$ and  $\mathbb{N}_p^{(n)}=p_{B_n}' + g_{n,n+1} \, p_{A_{n+1}}'= p_{B_{n}} + {\cal N}_p^{(\EB)}$.

\section*{Acknowledgements}
M.N.N. acknowledges the QuantERA project CLUSSTAR (8C2024003) of the MEYS of the Czech Republic. Project CLUSSTAR has received funding from the EU Horizon Programme under Grant Agreement No. 731473 and 101017733 (QuantERA). R.F. acknowledges the EU Horizon Programme under Grant Agreement No.101080173 (CLUSTEC) and the project CZ$.02.01.010022\_0080004649$ (QUEENTEC) of the EU and the Czech Ministry of Education, Youth and Sport. 


\bibliography{BiblioNonlocal.bib}


\end{document}